\documentclass[a4paper,12pt,eqno]{article}
\usepackage{theorem}
\usepackage{latexsym,amssymb,amsfonts,amsmath}
\newcommand{\RM}{\mathbb{R}}
\newcommand{\ZM}{\mathbb{Z}}
\newcommand{\QM}{\mathbb{Q}}

\newcommand{\qp}{\QM_p}
\newcommand{\zp}{\ZM_p}
\newcommand{\cp}{T^{(p)}}
\newcommand{\cpm}{T^{(p)} _M}

\newtheorem{thm}{Theorem}[section]
\newtheorem{lem}[thm]{Lemma}
\newtheorem{cor}[thm]{Corollary}
\newtheorem{pro}[thm]{Proposition}

\title{{\Large {\bf Continuous-time quantum walks \\ on ultrametric spaces}}}
\author{
{\small Norio KONNO}\\
{\scriptsize Department of Applied Mathematics, 
Faculty of Engineering, 
Yokohama National University}\\
{\scriptsize Hodogaya, Yokohama 240-8501, Japan}\\
{\scriptsize konno@ynu.ac.jp}\\
}
\date{\empty }
\begin{document}
\maketitle
\begin{abstract}
We introduce a continuous-time quantum walk on an ultrametric space corresponding to the set of $p$-adic integers and compute its time-averaged probability distribution. It is shown that localization occurs for any location of the ultrametric space for the walk. This result presents a striking contrast to the classical random walk case. Moreover we clarify a difference between the ultrametric space and other graphs, such as cycle graph, line, hypercube and complete graph, for the localization of the quantum case. Our quantum walk may be useful for a quantum search algorithm on a tree-like hierarchical structure.
\end{abstract}

\begin{small}
\end{small}

\section{Introduction}
Continuous-time classical random walks on ultrametric spaces have been investigated by some groups to describe the relaxation processes in complex systems, such as glasses, clusters and proteins \cite{ogi,ave99,pri,ave02,ave03,nec,ave04,luk}. On the other hand, a continuous-time quantum walk, which is a quantum version of the classical random walk, has been widely studied by various researchers for some important graphs, such as cycle graph, line, hypercube and complete graph, with the hope of finding a new quantum algorithmic technique \cite{chi02,chi03,ahm,ada,ben,inu,kon05,sol,fed,jaf}. However any result is not known for a quantum walk on the ultrametric space. This is a motivation for our present work. Excellent surveys of quantum walks are given in Refs. \cite{kem,tre}.

Let $X$ be a set and let $\rho$ be a metric on $X.$ The pair $(X, \rho)$ is called a metric space. Moreover if $\rho$ has the strong triangle inequality: 
\begin{eqnarray*}
\rho (x,z) \le \max ( \rho (x,y), \rho (y,z) ),
\end{eqnarray*}
for any $x,y,z \in X$, then $\rho$ is said to be an ultrametric. A set endowed with an ultrametric is called an ultrametric space. Let $p$ be a prime number and $\qp$ be the field of $p$-adic numbers. $\zp$ denotes the set of $p$-adic integers. Then $\zp$ is a subring of $\qp$ and $\zp = \{ x \in \qp : |x|_p \le 1 \},$ where $|\cdot|_p$ is the $p$-adic absolute value. Remark that the metric $\rho_p (x,y) = |x-y|_p$ is an ultrametric. Every $x \in \zp$ can be expanded in the following way: 
\begin{eqnarray*}
x = y_0 + y_1 p + y_2 p^2 + \cdots + y_n p^n + \cdots,  
\end{eqnarray*}
where $y_n \in \{0,1, \ldots, p-1 \}$ for any $n$. See Ref. \cite{khr} for more details. It is important that $\zp$ can be represented by the bottom infinite regular Cayley tree of degree $p$, $\cp$, in which every branch at each level splits into $p$ other branches. So we consider a continuous-time quantum walk on the bottom of $\cp$ at level (or depth) $M$, denoted by $\cpm$.

We calculate the probability distribution of the quantum walk on $\cpm$. From this, we obtain the time-averaged probability distribution on $\cpm$ and then take a limit as $M \to \infty$. On the other hand, first we take the limit as $M \to \infty$ and then we derive the time-averaged probability distribution. It is shown that the first limit result coincides with the second one, that is, time-averaged operation and $M$-limit one are commutative. By using the results, we clarify a difference between classical and quantum cases. A striking difference is that localization occurs at any site for a wide class of quantum walks. Furthermore we compare the results of ultrametric space with those of other graphs, such as cyclic graph, line, hypercube and complete graph in the quantum case. Some applications of $p$-adic analysis in physics and biology were reported \cite{khr}. Social network models based on the ultrametric distance were presented and studied \cite{wattsa,wattsb}. The disease spreading on a hierarchical metapopulation model was investigated \cite{wattsc}. Our quantum walk on the ultrametric space may be useful for designing a new quantum search algorithm on the graph with a hierarchical structure. 

This paper is organized as follows. Section 2 is devoted to the definition of a continuous-time quantum walk on $\cpm$. Our results are shown in Sec. 3. We review corresponding results for continuous-time classical random walks in Sec. 4. Section 5 contains results on other graphs for the quantum case. Conclusion is given in Sec. 6.

\section{Definition of Quantum Walk}
For $M \ge 0$ and $x \in \zp$, the closed $p$-adic ball $B_M ^{(p)} (x)$ of radius $p^{-M}$ with center $x$ is defined by
\begin{eqnarray*}
B_M ^{(p)} (x) = \{ y \in \zp : \rho_p (x,y) \le p^{-M} \}.
\end{eqnarray*}
Each ball $B_M ^{(p)} (x)$ of radius $p^{-M}$ can be represented as a finite union of disjoint balls $B_{M+1} ^{(p)} (x_m)$ of radius $p^{-{(M+1)}}$:
\begin{eqnarray*}
B_M ^{(p)} (x) = \bigcup_{m=0} ^{p-1} B_{M+1} ^{(p)} (x_m),
\end{eqnarray*}
for suitable $x_0, x_1, \ldots, x_{p-1} \in \zp,$ so we have 
\begin{eqnarray*}
\zp = \bigcup_{m=0} ^{p^M -1} B_{M} ^{(p)} (m).
\end{eqnarray*}
When $p=3$, 
\begin{eqnarray*}
\zp = \bigcup_{m=0} ^{3^1 -1} B_{1} ^{(3)} (m) = \bigcup_{k=0} ^{3^2 -1} B_{2} ^{(3)} (k),
\end{eqnarray*}
and

\begin{align*}
B_{1} ^{(3)} (0) 
&= B_{2} ^{(3)} (0) \cup B_{2} ^{(3)} (3) \cup B_{2} ^{(3)} (6), \\ 
B_{1} ^{(3)} (1) 
&= B_{2} ^{(3)} (1) \cup B_{2} ^{(3)} (4) \cup B_{2} ^{(3)} (7), \\ 
B_{1} ^{(3)} (2) 
&= B_{2} ^{(3)} (2) \cup B_{2} ^{(3)} (5) \cup B_{2} ^{(3)} (8).
\end{align*}

We define the distance between two balls $B_1$ and $B_2$ as 
\begin{eqnarray*}
\rho_p (B_1, B_2) = \inf \{ \rho_p (x,y) : x \in B_1, y \in B_2 \}.
\label{eqn:dis}
\end{eqnarray*}

We observe that end points of a regular Cayley tree with $p$ degree at level $M$ may be represented as a set of disconnected balls $\{ B_M ^{(p)} (0), B_M ^{(p)} (1), \ldots, B_M ^{(p)} (p^M-1) \}$ with radius $p^{-{M}}$ covering $\zp.$ So we can consider $p^M$ balls $\{ B_M ^{(p)} (0), B_M ^{(p)} (1), \ldots, B_M ^{(p)} (p^M-1) \}$ as $p^M$ points $\{0,1, \ldots , p^M-1 \}$, denoted by $\cpm.$

As in the case of classical random walk, let $\psi_M ^{(p)} (i,j)$ be amplitude of a jump from the ball $B_M ^{(p)} (i)$ to the ball $B_M ^{(p)} (j)$ separated by $p$-adic distance $p^{-(M-k)}.$ Here we put $\epsilon_k = \epsilon_M ^{(p)} (k) = \psi_M ^{(p)} (i,j)$. Remark that $\psi_M ^{(p)} (i,j) = \psi_M ^{(p)} (j,i).$

We consider $p=3$ and $M=2$. In this case, $\epsilon_k$ is given by
\begin{align*}
\epsilon_1 
&= 
\psi_2 ^{(3)} (0,3) = \psi_2 ^{(3)} (0,6) =  \psi_2 ^{(3)} (3,0) 
= \psi_2 ^{(3)} (3,6) 
\\
&= \psi_2 ^{(3)} (6,0) = \psi_2 ^{(3)} (6,3) = 
\psi_2 ^{(3)} (1,4) = \psi_2 ^{(3)} (1,7) = \cdots ,
\\
\epsilon_2 
&=
\psi_2 ^{(3)} (0,1) = \psi_2 ^{(3)} (0,4) =  \cdots = \psi_2 ^{(3)} (0,8) 
\\
&= 
\psi_2 ^{(3)} (3,1) = \psi_2 ^{(3)} (3,4) =  \cdots = \psi_2 ^{(3)} (3,8) = \cdots ,
\\
&\cdots \cdots . 
\end{align*}

Let $I_n$ and $J_n$ denote the $n \times n$ identity matrix and the all-one $n \times n$ matrix. We define a $p^M \times p^M$ symmetric matrix $H_M ^{(p)}$ on $\cpm$ which is treated as the Hamiltonian of the quantum system as follows: for any $M = 1,2, \ldots ,$
\begin{align*}
H_{M+1} ^{(p)} &= I_p \otimes H_M ^{(p)} + (J_p - I_p) \otimes \epsilon_{_{M+1}} I_{p^M}, \\
H_1 ^{(p)} &= \epsilon_0 I_p + \epsilon_1 (J_p - I_p), 
\label{eqn:hami}
\end{align*}
where $\epsilon_j \in \RM \> (j=0,1,2, \ldots )$ and $\RM$ is the set of real numbers.

The evolution of continuous-time quantum walk on $\cpm$ is governed by the following unitary matrix:
\begin{eqnarray*}
U^{(p)} _M (t) = e^{it H_M ^{(p)}}.
\label{eqn:unitary}
\end{eqnarray*}
The amplitude wave function at time $t$, $| \Psi_M ^{(p)} (t) \rangle $, is defined by 
\begin{eqnarray*}
| \Psi_M ^{(p)} (t) \rangle = U_M ^{(p)} (t) | \Psi_M ^{(p)} (0) \rangle . 
\label{eqn:evolution}
\end{eqnarray*}
In this paper we take $| \Psi_M ^{(p)} (0) \rangle = [1,0,0, \ldots, 0]^T$ as an initial state, where $T$ denotes the transposed operator.

The $(n+1)$-th coordinate of $| \Psi_M ^{(p)} (t) \rangle$ is denoted by $| \Psi_M ^{(p)} (n,t) \rangle$ which is the amplitude wave function at site $n$ at time $t$ for $n=0,1, \ldots , p^M -1.$ The probability finding the walker is at site $n$ at time $t$ on $\cpm$ is given by
\begin{eqnarray*}
P_N ^{(p)} (n,t) = \langle \Psi_N ^{(p)} (n,t) | \Psi_N ^{(p)} (n,t) \rangle .
\label{eqn:prob}
\end{eqnarray*}

\section{Our Results}
Let $\{ \eta_m : m=0,1, \ldots, M \}$ be defined by 
\begin{eqnarray*}
\eta_m = 
\left\{ 
\begin{array}{ll}
\epsilon_0 - \epsilon_1
& 
\qquad \hbox{if} \>\> m=0, 
\\
\epsilon_0 + (p-1) \sum_{k=1} ^m p^{k-1} \epsilon_k - p^m \epsilon_{m+1} 
&
\qquad \hbox{if} \>\> m=1, 2, \ldots, M-1, 
\\
0
& 
\qquad \hbox{if} \>\> m=M.
\end{array} \right.
\label{eqn:etam}
\end{eqnarray*}

Direct computation yields that $\{ \eta_m : m=0,1, \ldots, M \}$ is the set of eigenvalues of $H_M ^{(p)}$ and eigenvectors for $\eta_m$ are given in the following way, see Ref. \cite{ogi} for $p=2$ case. Let $\omega = \omega_p = \exp (2 \pi i /p)$. Put 
\begin{eqnarray*}
0_n = [\overbrace{0,0, \ldots , 0}^{n}], \quad 1_n = [\overbrace{1,1, \ldots , 1}^{n}].
\end{eqnarray*}
The $p^{M-1}(p-1)$ eigenvectors for $\eta_0$ are 
\begin{align*}
u_0 (1) 
&= [1, \omega, \omega^2, \ldots, \omega^{p-1}, 0_{p^M-p}]^T,
\\
u_0 (2) 
&= [1, \omega^2, \omega^4, \ldots, \omega^{2(p-1)}, 0_{p^M-p}]^T,
\\
&\ldots
\\
u_0 (p-1) 
&= [1, \omega^{(p-1)}, \omega^{2(p-1)}, \ldots, \omega^{(p-1)^2}, 0_{p^M-p}]^T,
\\
u_0 (p) 
&= [0_p, 1, \omega, \omega^2, \ldots, \omega^{p-1}, 0_{p^M-2p}]^T,
\\
u_0 (p+1) 
&= [0_p, 1, \omega^2, \omega^4, \ldots, \omega^{2(p-1)}, 0_{p^M-2p}]^T,
\\
&\ldots
\\
u_0 (2(p-1)) 
&= [0_p, 1, \omega^{(p-1)}, \omega^{2(p-1)}, \ldots, \omega^{(p-1)^2}, 0_{p^M-2p}]^T,
\\
&\ldots
\\
u_0 (p^{M-1}(p-1)) 
&= [0_{p^M-p},1, \omega^{(p-1)}, \omega^{2(p-1)}, \ldots, \omega^{(p-1)^2}]^T.
\end{align*}
The $p^{M-2}(p-1)$ eigenvectors for $\eta_1$ are 
\begin{align*}
u_1 (1) 
&= [1_p, \omega 1_p, \omega^2 1_p, \ldots, \omega^{p-1} 1_p, 0_{p^M-p^2}]^T,
\\
u_1 (2) 
&= [1_p, \omega^2 1_p, \omega^4 1_p, \ldots, \omega^{2(p-1)} 1_p, 0_{p^M-p^2}]^T,
\\
&\ldots
\\
u_1 (p-1) 
&= [1_p, \omega^{(p-1)} 1_p, \omega^{2(p-1)} 1_p, \ldots, \omega^{(p-1)^2} 1_p, 0_{p^M-p^2}]^T,
\\
u_1 (p) 
&= [0_{p^2}, 1_p, \omega 1_p, \omega^2 1_p, \ldots, \omega^{p-1} 1_p, 0_{p^M-2p^2}]^T,
\\
u_1 (p+1) 
&= [0_{p^2}, 1_p, \omega^2 1_p, \omega^4 1_p, \ldots, \omega^{2(p-1)} 1_p, 0_{p^M-2p^2}]^T,
\\
&\ldots
\\
u_1 (2(p-1)) 
&= [0_{p^2}, 1_p, \omega^{(p-1)} 1_p, \omega^{2(p-1)} 1_p, \ldots, \omega^{(p-1)^2} 1_p, 0_{p^M-p^2}]^T,
\\
&\ldots
\\
u_1 (p^{M-2} (p-1)) 
&= [0_{p^M-p^2},1_p, \omega^{(p-1)} 1_p, \omega^{2(p-1)} 1_p, \ldots, \omega^{(p-1)^2} 1_p]^T.
\end{align*}
We have $p^{M-(k+1)}(p-1)$ eigenvectors for $\eta_{k} \> (k=2, 3, \ldots, M-1)$ similarly. So the $p-1$ eigenvectors for $\eta_{M-1}$ are 
\begin{align*}
u_{M-1} (1) 
&= [1_{p^{M-1}}, \omega 1_{p^{M-1}}, \omega^2 1_{p^{M-1}}, \ldots, \omega^{p-1} 1_{p^{M-1}}]^T,
\\
u_{M-1} (2) 
&= [1_{p^{M-1}}, \omega^2 1_{p^{M-1}}, \omega^4 1_{p^{M-1}}, \ldots, \omega^{2(p-1)} 1_{p^{M-1}}]^T,
\\
&\ldots
\\
u_{M-1} (p-1) 
&= [1_{p^{M-1}}, \omega^{(p-1)} 1_{p^{M-1}}, \omega^{2(p-1)} 1_{p^{M-1}}, \ldots, \omega^{(p-1)^2} 1_{p^{M-1}}]^T.
\end{align*}
Finally the eigenvector for $\eta_{M}$ is only $[1_{p^M}]^T.$

Throughout this paper, we assume that 
\begin{eqnarray*}
0 < \epsilon_M < \epsilon_{M-1} < \cdots < \epsilon_2 < \epsilon_1.
\end{eqnarray*}
The assumption implies that our quantum walk class does not belong to a class of continuous-time quantum walks given by the adjacency matrix of the graph on which the walk is defined, see Ref. \cite{jaf}.

Remark that the diagonal component $\epsilon_0$ is an irrelevant phase factor in the wave evolution. So the probability distribution at time $t$, $\{ P_M ^{(p)} (n,t) : n=0, 1, \ldots , p^M-1 \}$, does not depend on $\epsilon_0$ and we put 
\begin{eqnarray*}
\epsilon_0 = - (p-1) \sum_{k=1} ^M p^{k-1} \epsilon_k < 0, 
\end{eqnarray*}
as in the classical case. We should note that $\epsilon_0 < 0 < \epsilon_M < \epsilon_{M-1} < \cdots < \epsilon_2 < \epsilon_1$ is equivalent to $\eta_0 < \eta_1 < \cdots < \eta_{M-1} < \eta_{M} =0$.

Let $V_{k} ^{(p)} = \{ p^{k-1}, p^{k-1} +1, \ldots, p^{k}-1 \} \> (k=1, 2, \ldots, M)$ and $V_0 ^{(p)} = \{0\}$. Remark that $\bigcup_{k=0} ^M V_{k} ^{(p)} = \{0,1, \ldots, p^M-1 \}$ and $|V_k ^{(p)}| = (p-1)p^{k-1}$ for $k=1,2, \ldots, M$, where $|A|$ is the number of elements in a set $A$. By using the eigenvectors, we obtain the amplitude of the quantum walk on $\cpm$:
\begin{lem}
\label{lem31}
\begin{eqnarray*}
| \Psi_M ^{(p)} (n,t) \rangle = 
\left\{ 
\begin{array}{ll}
\displaystyle{(p-1) \sum_{m=0} ^{M-1} p^{-(m+1)} e^{it \eta_m} + p^{-M}}
& 
\\
\qquad \qquad \hbox{if} \>\> n \in V_0 ^{(p)}, 
&
\\
\displaystyle{- p^{-k} e^{it \eta_{k-1}} + (p-1) \sum_{m=k} ^{M-1} p^{-(m+1)} e^{it \eta_m} + p^{-M}} 
& 
\\
\qquad \qquad \hbox{if} \>\> n \in V_k ^{(p)} \>\> (k=1,2, \ldots, M-1), 
&
\\
\displaystyle{p^{-M}( - e^{it \eta_{M-1}} + 1 )}
& 
\\
\qquad \qquad \hbox{if} \>\> n \in V_M ^{(p)}.
\end{array} \right.
\end{eqnarray*}
\end{lem}

The definition of $P_M ^{(p)} (n,t)$ implies
\begin{pro}
\label{pro32}
\begin{eqnarray*}
P_M ^{(p)} (n,t) = 
\left\{ 
\begin{array}{ll}
\displaystyle{ 
\left\{ 
(p-1) \sum_{m=0} ^{M-1} p^{-(m+1)} \cos (t \eta_m) + p^{-M}
\right\}^2
}
&
\\
\qquad \qquad \qquad \qquad \qquad 
\displaystyle{ 
+
(p-1)^2
\left\{ 
\sum_{m=0} ^{M-1} p^{-(m+1)} \sin (t \eta_m)
\right\}^2
}
& 
\\
\qquad \qquad \hbox{if} \>\> n \in V_0 ^{(p)}, 
&
\\
\displaystyle{
\left\{ 
- p^{-k} \cos (t \eta_{k-1}) 
+ (p-1) \sum_{m=k} ^{M-1} p^{-(m+1)} \cos (t \eta_m) 
+ p^{-M}
\right\}^2
}
&
\\
\qquad \qquad
\displaystyle{
+
\left\{ 
- p^{-k} \sin (t \eta_{k-1}) 
+ (p-1) \sum_{m=k} ^{M-1} p^{-(m+1)} \sin (t \eta_m) 
\right\}^2
} 
& 
\\
\qquad \qquad \hbox{if} \>\> n \in V_k ^{(p)} \>\> (k=1,2, \ldots, M-1), 
&
\\
\displaystyle{
2 p^{-2M} (1- \cos (t \eta_{_{M-1}}))
}
& 
\\
\qquad \qquad \hbox{if} \>\> n \in V_M ^{(p)},
\end{array} \right.
\end{eqnarray*}
\end{pro}

If $n=0$, then $P_M ^{(p)} (0,t)$ is the return probability of the walk on $\cpm.$ For $p=3$ and $M=2$ case, we obtain 
\begin{eqnarray*}
P_2 ^{(3)} (n,t) = 
\left\{ 
\begin{array}{ll}
\left\{ 41 + 24 \cos (t (\eta_0 - \eta_1)) + 12 \cos (t \eta_0) + 4 \cos (t \eta_1) \right\}/3^4
& 
\\
\qquad \qquad \qquad \hbox{if} \>\> n=0, 
&
\\
\left\{ 14 - 12 \cos (t (\eta_0 - \eta_1)) - 6 \cos (t \eta_0) + 4 \cos (t \eta_1) \right\}/3^4
&
\\
\qquad \qquad \qquad \hbox{if} \>\> n=1,2, 
\\
2 \left\{ 1 -  \cos (t \eta_1) \right\}/3^4
&
\\
\qquad \qquad \qquad \hbox{if} \>\> n=3,4, \ldots ,8.
\end{array} \right.
\end{eqnarray*}
In general, $P_{M} ^{(p)}(n,t)$ does not converge as $t \to \infty$ for any fixed $n$. So we introduce a time-averaged distribution of $P_{M} ^{(p)}(n,t)$ as follows:
\begin{eqnarray}
\bar{P}_{M} ^{(p)} (n)=\lim_{t \rightarrow \infty} \frac{1}{t} \int_{0}^{t} P_{M} ^{(p)}(n,s) ds ,
\label{eqn:averageconti}
\end{eqnarray}
if the right-hand side of Eq. (\ref{eqn:averageconti}) exists. Then Proposition {\rmfamily \ref{pro32}} gives

\begin{thm}
\label{thm33}
\begin{eqnarray*}
\bar{P}_{M} ^{(p)} (n) = 
\left\{ 
\begin{array}{ll}
\displaystyle{ 
{p-1 \over p+1} +{2 \over (p+1)p^{2M}}
}
& 
\qquad \hbox{if} \>\> n \in V_0 ^{(p)}, 
\\
\displaystyle{
{2 \over p+1} \left( {1 \over p^{2k-1}} +{1 \over p^{2M}} \right) 
} 
& 
\qquad \hbox{if} \>\> n \in V_k ^{(p)} \>\> (k=1,2, \ldots, M-1), 
\\
\displaystyle{
{2 \over p^{2M}}
}
& 
\qquad \hbox{if} \>\> n \in V_M ^{(p)}.
\end{array} \right.
\end{eqnarray*}
\end{thm}
It is interesting to note that $\bar{P}_{M} ^{(p)} (n)$ does not depend on $\{ \epsilon_k : k=0,1, \ldots, M \}.$ In the case of $p=3$ and $M=2$, we have
\begin{eqnarray*}
\bar{P}_{2} ^{(3)} (0) = {41 \over 3^4}, \> 
\bar{P}_{2} ^{(3)} (1) = \bar{P}_{2} ^{(3)} (2) =  {14 \over 3^4}, \> 
\bar{P}_{2} ^{(3)} (3) = \cdots = \bar{P}_{2} ^{(3)} (8) =  {2 \over 3^4}.
\end{eqnarray*}
The following result is immediate from Theorem {\rmfamily \ref{thm33}}.

\begin{cor}
\label{cor34}
\begin{eqnarray*}
\lim_{M \to \infty} \bar{P}_{M} ^{(p)} (n) = 
\left\{ 
\begin{array}{ll}
\displaystyle{ 
{p-1 \over p+1}
}
& 
\qquad  \hbox{if} \>\> n \in V_0 ^{(p)}, 
\\
\displaystyle{
{2 \over (p+1) p^{2k-1}}  
} 
& 
\qquad \hbox{if} \>\> n \in V_k ^{(p)} \>\> (k=1,2, \ldots ).
\end{array} \right.
\end{eqnarray*}
\end{cor}

Here we consider the mean distance from $0$ at time $t$ defined by
\begin{eqnarray*}
d_{M} ^{(p)} (t) = \sum_{k=1} ^M \sum_{n \in V_k ^{(p)}} p^{-(M-k)} P_{M} ^{(p)} (n,t).
\end{eqnarray*}
Then its time-averaged mean distance is given by
\begin{eqnarray*}
\bar{d}_{M} ^{(p)} = \sum_{k=1} ^M \sum_{n \in V_k ^{(p)}} p^{-(M-k)} \bar{P}_{M} ^{(p)} (n).
\end{eqnarray*}
From Theorem {\rmfamily \ref{thm33}}, we get
\begin{eqnarray*}
\bar{d}_{M} ^{(p)} = {2(p-1)(M-1) \over (p+1)p^M} 
+ { 2 \left[ \left\{ (p-1)(p+1)^2+1 \right\} p^{2M-2} - 1 \right] \over (p+1)^2 p^{3M-1}}.
\end{eqnarray*}
This gives
\begin{eqnarray*}
\lim_{M \to \infty} {p^M \over M} \bar{d}_{M} ^{(p)} = {2(p-1) \over p+1}. 
\end{eqnarray*}

Next we take a limit as $M \to \infty$ first. We define $P_{\infty} ^{(p)} (n,t)$ by $P_{\infty} ^{(p)} (n,t)= \lim_{M \to \infty} P_M ^{(p)} (n,t)$, if the right-hand side of the equation exists. By Proposition {\rmfamily \ref{pro32}}, we obtain

\begin{pro}
\label{pro35}
\begin{eqnarray*}
P_{\infty} ^{(p)} (n,t) = 
\left\{ 
\begin{array}{ll}
\displaystyle{ 
(p-1)^2 \left[ \left\{ 
\sum_{m=0} ^{\infty} p^{-(m+1)} \cos (t \eta_m)
\right\}^2 \right.
}
&
\\
\qquad \qquad \qquad \qquad \qquad \qquad \qquad
\displaystyle{ 
\left.
+
\left\{ 
\sum_{m=0} ^{\infty} p^{-(m+1)} \sin (t \eta_m)
\right\}^2
\right]
}
& 
\\
\qquad \qquad \hbox{if} \>\> n \in V_0 ^{(p)}, 
&
\\
\displaystyle{
\left\{ 
- p^{-k} \cos (t \eta_{k-1}) 
+ (p-1) \sum_{m=k} ^{\infty} p^{-(m+1)} \cos (t \eta_m) 
\right\}^2
}
&
\\
\qquad \qquad
\displaystyle{
+
\left\{ 
- p^{-k} \sin (t \eta_{k-1}) 
+ (p-1) \sum_{m=k} ^{\infty} p^{-(m+1)} \sin (t \eta_m) 
\right\}^2
} 
& 
\\
\qquad \qquad \hbox{if} \>\> n \in V_k ^{(p)} \>\> (k=1,2, \ldots ).
\end{array} \right.
\end{eqnarray*}
\end{pro}
In a similar way, a time-averaged distribution $P_{\infty} ^{(p)} (n,t)$ is given by
\begin{eqnarray}
\bar{P}_{\infty} ^{(p)} (n) = \lim_{t \rightarrow \infty} \frac{1}{t} \int_{0}^{t} P_{\infty} ^{(p)} (n,s) ds ,
\label{eqn:averagecontiinfty}
\end{eqnarray}
if the right-hand side of Eq. (\ref{eqn:averagecontiinfty}) exists. Combining Theorem {\rmfamily \ref{thm33}} with Proposition {\rmfamily \ref{pro35}} yields

\begin{cor}
\label{cor36}
For any $n=0,1,2, \ldots$, we have
\begin{eqnarray*}
\bar{P}_{\infty} ^{(p)} (n) = \lim_{M \to \infty} \bar{P}_{M} ^{(p)} (n) >0.
\end{eqnarray*}
\end{cor}
Furthermore, 
\begin{eqnarray*}
\lim_{p \to \infty} \bar{P}_{\infty} ^{(p)} (n) = \delta_0 (n),
\end{eqnarray*}
where $\delta_m (n) = 1,$ if $n=m, \> =0,$ if $n \not=m.$ 

In general, we say that localization occurs at site $n$ if the time-averaged probability at the site is positive. Therefore the localization occurs at any site for both any finite $M$ and $M \to \infty$ limit cases.

\section{Classical Case}
In this section we review three classical examples and clarify a difference between classical and quantum walks. Let $P_{c} ^{(p)} (n,t)$ be the probability that a classical random walker starting from $0$ is located at site $n$ at time $t$ on $\zp$.

In the case of a linear landscape, the transition rate on $\cpm$ has the form 
\begin{eqnarray*}
\epsilon_k = w_0 \> p^{- (1+ \alpha) (k-M)},
\end{eqnarray*}
for $w_0 >0$ and $\alpha >0$. Then Avetisov {\it et al.} \cite{ave02} showed a power decay law in $M \to \infty$ limit: 
\begin{eqnarray*}
P_{c} ^{(p)} (0,t) \sim {1 \over t^{1/\alpha}} \quad (t \to \infty),
\end{eqnarray*}
where $f(t) \sim g(t) \> (t \to \infty)$ means there exist positive constants $C_1$ and $C_2$ such that 
\begin{eqnarray*}
C_1 \le \liminf_{t \to \infty} { f(t) \over g(t)} \le \limsup_{t \to \infty} {f(t) \over g(t)} \le C_2 .
\end{eqnarray*}

For a logarithmic landscape case, the transition rate on $\cpm$ is 
\begin{eqnarray*}
\epsilon_k = w_0 \> p^{-(k-M)} {1 \over ( \log(1+p^{-(k-M)}))^{\alpha} },
\end{eqnarray*}
for $w_0 >0$ and $\alpha >1$. The following stretched exponential decay law (the Kohlrausch-Williams-Watts law) was proved by Avetisov {\it et al.} \cite{ave03} in $M \to \infty$ limit: 
\begin{eqnarray*}
\log (P_{c} ^{(p)} (0,t)) \sim - t^{1/\alpha} \quad (t \to \infty).
\end{eqnarray*}

In the case of an exponential landscape, the transition rate on $\cpm$ has the form 
\begin{eqnarray*}
\epsilon_k = w_0 \> p^{-(k-M)} \exp ( - \alpha p^{(k-M)} ) ,
\end{eqnarray*}
for $w_0 >0$ and $\alpha >0$.
Then Avetisov {\it et al.} \cite{ave03} obtained a logarithmic decay law taking a limit as $M \to \infty$: 
\begin{eqnarray*}
P_{c} ^{(p)} (0,t) \sim {1 \over \log t} \quad (t \to \infty).
\end{eqnarray*}
The above three facts imply that the localization does not occur at position $0$. In contrast to the classical case, the localization occurs at any position for our quantum case.

\section{Quantum Case for Other Graphs}
We consider the time-averaged probability distribution for continuous-time quantum walk starting from a site on other graphs, such as cycle graph, line, hypercube and complete graph. Then the Hamiltonian of the walk is given by the adjacency matrix of the graph.

In the case of a cycle graph $C_N$ with $N$ sites, we have obtained the following result \cite{inu}: 
\begin{eqnarray*}
\bar{P}_{N}(n)= {1 \over N} + {2 R_N (n) \over N^2} ,
\end{eqnarray*}
for any $n=0,1, \ldots, N-1$, where 
\begin{eqnarray*}
R_N (n) =\left\{ 
\begin{array}{ll}
-1/2 & \qquad \hbox{if} \>\> N = \hbox{odd}, \quad \xi_{2n} \not= 0 \quad (\hbox{mod} \> 2 \pi), 
\\
-1 & \qquad \hbox{if} \>\> N = \hbox{even}, \quad \xi_{2n} \not= 0 \quad (\hbox{mod} \> 2 \pi), 
\\
\bar{N}  & \qquad \hbox{if} \>\> \xi_{2n} = 0 \quad (\hbox{mod} \> 2 \pi). 
\end{array} \right.
\end{eqnarray*}
Here $\xi_j = 2 \pi j/N, \> \bar{N}=[(N-1)/2],$ and $[x]$ is the smallest integer greater than $x$. When $N =$ odd (i.e., $\bar{N}=(N-1)/2$),
\begin{eqnarray*}
\bar{P}_N = \Bigl( {1 \over N}+{N-1 \over N^2}, 
\overbrace{{1 \over N} - {1 \over N^2}, \ldots , {1 \over N} - {1 \over N^2}}^{N-1} \Bigr),
\end{eqnarray*}
when $N =$ even (i.e., $\bar{N}=(N-2)/2$),
\begin{eqnarray*}
\bar{P}_N = \Bigl( {1 \over N}+{N-2 \over N^2}, 
\overbrace{{1 \over N} - {2 \over N^2}, \ldots , {1 \over N} - {2 \over N^2}}^{(N-2)/2}, {1 \over N}+{N-2 \over N^2}, 
\overbrace{{1 \over N} - {2 \over N^2}, \ldots , {1 \over N} - {2 \over N^2}}^{(N-2)/2} \Bigr).
\end{eqnarray*}
Then we have
\begin{eqnarray*}
\lim_{N \to \infty} \bar{P}_N (n) = 0,
\end{eqnarray*}
for any $n$.

In $\ZM$ case, the probability distribution $P(n,t)$ of the walk starting from the origin at location $n$ and time $t$ is given by  
\begin{eqnarray*}
P (n,t) = J_{n} ^2 (t) , 
\end{eqnarray*}
where $J_n (t)$ is the Bessel function of the first kind of order $n$, see Ref. \cite{kon05}, for example. The asymptotic behavior of $J_n (t)$ at infinity is as follows:
\begin{eqnarray*}
J_n (t) = \sqrt{{2 \over \pi t}} \left( \cos(t- \theta (n)) - \sin (t - \theta (n)) {4 n^2 -1 \over 8t} + O(t^{-2}) \right), \quad t \to \infty,
\end{eqnarray*}
where $\theta (n) = (2 n + 1) \pi/4$, see page 195 in Ref. \cite{wat}. From this fact and $|J_n (t)| \le 1$ for any $t$ and $n$ (see page 31 in Ref. \cite{wat}), we obtain
\begin{eqnarray*}
\bar{P} (n) = \lim_{t \to \infty} {1 \over t} \int_0 ^{t}  J_{n} ^2 (s) \> ds = 0,
\end{eqnarray*}
for any $n \in \ZM.$ 

Next we consider the quantum walk starting from a site, denoted by $0$, on a hypercube with $2^N$ sites, $W_N$. Then the probability finding the walker at site $n$ at time $t$ is 
\begin{eqnarray*}
P_N (n,t) = \cos (t/N)^{2(N-k)} \sin(t/N) ^{2k}
\qquad \hbox{if} \>\> n \in V_k \quad (0 \le k \le N), 
\end{eqnarray*}
where $V_k = \{x \in W_N : ||x||=k \}$ and $|| x ||$ is the path length from $0$ to $x$ in $W_N$. A derivation is shown in Ref. \cite{jaf}. From the result we see that
\begin{eqnarray*}
\bar{P}_N (n) = {1 \over 2^{2N}} {N \choose k}^{-1} {2k \choose k} {2(N-k) \choose N-k}, 
\end{eqnarray*}
if $n \in V_k \> (0 \le k \le N).$ For example, when $N=4$, 
\begin{eqnarray*}
\bar{P}_4 (n) = \left\{ 
\begin{array}{ll}
35/128
&
\qquad \hbox{if} \>\> n \in V_0 \cup V_{4},
\\
5/128
& 
\qquad \hbox{if} \>\> n \in V_1 \cup V_{3},
\\
3/128
& 
\qquad \hbox{if} \>\> n \in V_2.
\end{array} \right.
\end{eqnarray*}
In general $N$, we have
\begin{eqnarray*}
\lim_{N \to \infty} \bar{P}_N (n) = 0,
\end{eqnarray*}
for any $n =0,1, \ldots $. Moreover if we let 
\begin{eqnarray*}
\bar{P}_N (V_k) = \sum_{n \in V_k} \bar{P}_N (n),
\end{eqnarray*}
for $k=0,1, \ldots, N,$ then we get
\begin{eqnarray*}
\bar{P}_N (V_k) = {1 \over 2^{2N}} {2k \choose k} {2(N-k) \choose N-k},
\end{eqnarray*}
since $|V_k|= N!/(N-k)!k!.$ It is interesting to note that this probability corresponds to a well-known arcsin law for the classical random walk. 

Therefore, in three cases, $C_N$ (cycle graph) and $W_N$ ($N$-cube) as $N \to \infty$ and $\ZM$, the localization does not occur at any location.

In the case of the complete graph with $N$ sites, $K_N$, the Hamiltonian $H_N$ is defined by $H_N = I_N - N J_N$. Then eigenvalues of $H_N$ are $\eta_0 = 0, \eta_1 = \eta_2 = \cdots = \eta_{N-1} = -N$. The eigenvectors are given by the Vandermonde matrix, see Ref. \cite{ahm}. Then direct computation yields 
\begin{eqnarray*}
P_N (n,t) =\left\{ 
\begin{array}{ll}
\displaystyle{ {(N-1)^2 + 1 + 2(N-1) \cos (Nt) \over N^2}}
&
\qquad \hbox{if} \>\> n = 0, 
\\
\displaystyle{ {2 (1 -  \cos (Nt)) \over N^2}}
& 
\qquad \hbox{if} \>\> n = 1,2, \ldots, N-1. 
\end{array} \right.
\end{eqnarray*}
Therefore
\begin{eqnarray*}
\bar{P}_N (n) =\left\{ 
\begin{array}{ll}
\displaystyle{ {(N-1)^2+1 \over N^2}}
&
\qquad \hbox{if} \>\> n = 0, 
\\
\displaystyle{ {2 \over N^2}}
& 
\qquad \hbox{if} \>\> n = 1,2, \ldots, N-1. 
\end{array} \right.
\end{eqnarray*}
So we have 
\begin{eqnarray*}
\lim_{N \to \infty} \bar{P}_N (n) = \delta_0 (n),
\end{eqnarray*}
for any $n =0,1, \ldots .$ Therefore the localization occurs at only $n=0$ site. This corresponds to our case for $p \to \infty$.

\section{Conclusion}
We have derived the expression of the probability distribution of a continuous-time quantum walk on $\cpm$ corresponding to $\zp$ in the $M \to \infty$ limit. As a result, we obtained
\begin{eqnarray*}
\bar{P}_{\infty} ^{(p)} (n) = \lim_{M \to \infty} \bar{P}_{M} ^{(p)} (n) >0, \quad (n=0,1,2, \ldots),
\end{eqnarray*}
for a class of $\epsilon_k$ satisfying
\begin{eqnarray*}
\epsilon_0 <0< \epsilon_M < \epsilon_{M-1} < \cdots < \epsilon_2 < \epsilon_1.
\end{eqnarray*}
Therefore the localization occurs at any location. For $C_N$ (cycle graph) and $W_N$ ($N$-cube) as $N \to \infty$ and $\ZM$ cases, the localization does not happen at any location. For $K_N$ (complete graph) as $N \to \infty$ case, the localization occurs at only $0$ site. In three typical classical cases, the localization does not occur even at 0 site. We hope that this property of our quantum walk can be useful in a search problem on a tree-like hierarchical structure. 
\par
\
\par\noindent
{\bf Acknowledgments}
The author is grateful to Naoki Masuda, Shigeki Matsutani and Etsuo Segawa for valuable discussions and comments.

\begin{small}
\bibliographystyle{jplain}

\end{small}

\end{document}